# Surface states in AlGaN/GaN high electron mobility transistors: Quantitative profiles and dynamics of the surface Fermi level

Yury Turkulets,[1] and Ilan Shalish[1,a]

**AFFILIATIONS**

[1] Department of Electrical and Computer Engineering, Ben-Gurion University, Beer Sheva 8410501, Israel

[a] Author to whom correspondence should be addressed: shalish@bgu.ac.il

**ABSTRACT**

We present a method to obtain quantitative profiles of surface state charge density and monitor its dynamics under various stress conditions in high electron mobility transistor (HEMT) devices. The method employs an optical spectroscopy of the channel current at various bias conditions. We test the method on a classical AlGaN/GaN HEMT structure. To analyze the results, we propose a model, according to which the energy distribution of the surface charge density may be obtained from the derivative of the channel photocurrent. The proposed method is applied to fully fabricated transistors and can be measured under any device bias combination. This way, it is possible to explore the effect of device operating conditions on the surface state charge. This feature should be especially useful in studies of the various surface charge migration effects in nitride HEMTs. An important byproduct of the method is a quantitative assessment of the energy position of the surface Fermi level and its dynamics under various bias conditions.

Surface states are known to constitute the origin of the two-dimensional electron gas (2DEG) charge in the nitride high electron mobility transistor (HEMT).[1,2] However, the same surface states are also held responsible for several adverse effects on device performance. Under certain operational conditions, charge originating at the surface undesirably migrates to various locations in the transistor, thereby changing its characteristics.[3,4,5,6] This places a high value on methods to characterize HEMT surface states.

To date, several methods have been proposed for surface state characterization in HEMTs. Higashiwaki et al. used series of AlGaN layer thicknesses as well as a series of Al compositions to show that the density of surface states is generally low while their profile is distributed.[7] Photo-assisted capacitance-voltage (CV) has been utilized for the characterization of surface states in insulator/GaN and insulator/AlGaN/GaN structures.[8,9,10] However, since this method requires a sweep of the gate voltage to positive values, its use may be challenging, and limited, in the case of Schottky-gate HEMT structures. A different approach was taken by Klein et al.[11] They illuminated the transistor at a specific photon energy, releasing trapped charges from deep levels into the 2DEG, and measuring the consequent increase in the channel conductivity. This way they were able to identify several spectral features related to deep level in the resulting photocurrent spectrum, but did not provide distribution profiles of the related traps.

In this study, we present a simple method to measure the energetic distribution of charge trapped in surface states. We show how a set of surface charge profiles acquired under different gate stress[12] conditions reveal the dynamics of the surface trapping mechanisms. The same method also reveals the exact position of the Fermi level at the surface along with the total trapped charge, both as a function of the gate voltage.

A standard AlGaN/GaN HEMT structure was grown on c-plane sapphire by metal-organic chemical vapor deposition (MOCVD) with the following layer sequence: an AlN nucleation layer, 2 μm GaN, 11 nm $Al_{0.3}Ga_{0.7}N$, and 2 nm of GaN cap-layer. Shallow mesas were dry-etched in chlorine-based plasma. 100 nm of $Si_3N_4$ passivation layer was deposited using plasma-enhanced chemical vapor deposition (PECVD), followed by dry etch of contact pads and gate trenches. Source and drain Ohmic contacts made of Ti(30 nm)/Al(70 nm)/Ni(30 nm)/Au(100 nm) were deposited by e-beam thermal evaporation, followed by thermal anneal at 900 °C for 1 min in $N_2$ ambient. Gate Schottky Ni(30 nm)/Au(100 nm) contacts were deposited by e-beam thermal evaporation. All electro-optical measurements were carried out in a dark Faraday box under ambient conditions.[13] The transistor was illuminated from its top (gate) side using monochromatic light. The light source was a 300 W Xe short-arc lamp monochromitized using a double Newport Corp. MS257 monochromator and filtered by a set of order-sorting long-pass filters. The photon energy was stepped from 1.5 to 3.2 eV in equal steps of 5 meV. Since the emission time of the trapped electron is typically long and varies with the excitation energy[14] (we measured 90 sec at photon energy of 3.1 eV), the samples were illuminated at each photon energy point for 120 sec prior to starting the $I_{DS}$ measurements (carried out under the same illumination) to ensure that the slow charge de-trapping process has reached a steady state. Closed loop control was employed to maintain a constant photon flux throughout the spectral acquisition.[15] Electrical measurements were carried out using two Keithley 2400 source meters. The first of which was used to apply a constant drain-to-source bias ($V_{DS}$) of 0.1 V and measure the drain-to-source current ($I_{DS}$), while the other was used to apply a constant gate-to-source ($V_{GS}$) voltage. The IDS value was acquired by averaging of 100 consecutive measurements at each data point. Constant $V_{GS}$ bias was applied four hours before, as well as throughout, the spectral acquisition to facilitate a steady state of the trap population. Transconductance was obtained from the drain current, measured by sweeping the gate voltage at 10 mV steps in the dark.

**Figure 1** shows the band diagram of the transistor at zero gate bias. As long as the transistor is kept in the dark (Fig. 1a), the surface states above the Fermi level are empty, while their electrons are in the channel constituting the 2DEG. In contrast, the surface states below the Fermi level are full, maintaining their original charge. Under illumination at a certain photon energy, hv, surface charge originating at an energy hv below the conduction band minimum (CBM) is excited, and consequently, swept into the channel by the strong polarization built-in field in the AlGaN. This photo-emitted charge increases the sheet charge density in the 2DEG by the amount of $\delta n_s(hv)$. The amount of charge added to the 2DEG is proportional to the total charge located at the energy $E_{CBM}$-hv at the surface. Thus, the photocurrent spectrum obtained by scanning the photon energy from low to high values is proportional to the integral over the surface charge density profile. This means that the derivative of the photocurrent spectrum is proportional to the energy profile of the surface charge density.

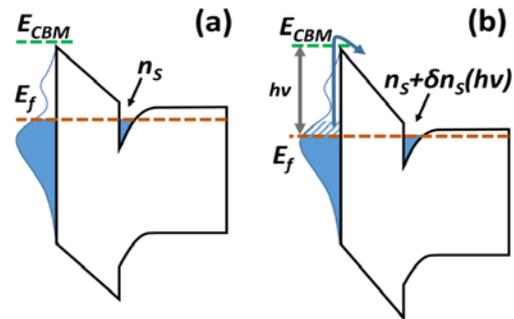

Fig. 1 Band diagram of the HEMT with an arbitrary distribution of surface states. (a) In the dark, the states above the Fermi level are empty, while those below are full. (b) Illumination by a certain photon energy, hv, releases trapped surface charge at an energy offset of hv below the CBM. Strong AlGaN built-in field pushes this charge into the 2DEG, thereby increasing its sheet charge by $\delta n_s$.

To analyze quantitatively the photo-induced changes in the photocurrent spectrum, one should take into account the concurrent effect of the gate bias. During the illumination, part of the trapped surface charge is photo-emitted over the AlGaN into the channel. This increases the drain current and simultaneously slightly decreases the built-in voltage in the AlGaN. As long as these changes are small compared to the values in the dark, we can write:

$$\Delta I_{DS}(hv) = g_m \Delta V_{GS}(hv) \quad (1)$$

where $g_m$ is the transconductance at the applied gate voltage, $\Delta I_{DS}(hv)$ is the photocurrent ($I_{ph}(hv)$), and $\Delta V_{GS}(hv)$ is the photo-induced change in the built-in voltage across the AlGaN layer. Since the AlGaN layer may be described as a parallel-plate capacitor, the photo-induced change in the built-in voltage is given by:

$$\Delta V_{GS}(hv) = \frac{Q(hv)}{C_{AlGaN}} \quad (2)$$

where $Q(hv)$ is the total areal charge photo-emitted over the AlGaN up to a specific photon energy, and $C_{AlGaN}$ – the AlGaN areal capacitance, given by $\varepsilon_0 \varepsilon_r \frac{1}{t_{AlGaN}}$, where $\varepsilon_0$ – absolute permittivity, $\varepsilon_r$ – relative permittivity, and $t_{AlGaN}$ is the thickness of the AlGaN layer. Substituting Eq. (2) into Eq. (1), we get:

$$Q(hv) = \frac{I_{ph}(hv)\varepsilon_0\varepsilon_r}{g_m t_{AlGaN}} \quad (3)$$

To calculate how much charge migrates over the AlGaN at each photon energy, we need to differentiate the total charge with respect to photon energy. This derivative is directly proportional to the charge density in the occupied surface states as a function of photon energy with respect to the CBM:

$$N_{OC}(E_{CBM} - hv) \propto \frac{1}{q}\frac{dQ(hv)}{dhv} = \frac{\varepsilon_0\varepsilon_r}{qg_m t_{AlGaN}}\frac{dI_{ph}(hv)}{dhv} \quad (4)$$

where q is the electron charge.

To quantitatively approximate the total charge density at the surface, the photon flux should be sufficient to excite most of the surface charge. However, in this case, charge migration will significantly change the photo-induced voltage across the AlGaN, $\Delta V_{GS}(h\nu)$, causing a major change in the transconductance during the measurement. Fortunately, the effect of this change can be easily taken into account by measuring the transconductance characteristics in the dark, and using the transconductance as a function of $\Delta V_{GS}(h\nu)$ instead of a constant value.[16]

The surface charge profiles we thus obtain drop to zero around the Fermi level. This enables us to extract the surface Fermi level position from the charge profiles. The transition between the occupied and unoccupied states is described by Fermi-Dirac distribution function. Thus, the surface charge density may be described by the following expression:

$$N_{OC}(E_{CBM} - h\nu) = N_{SS}(E_{CBM} - h\nu)\left(1 - \frac{1}{1 + exp\left(\frac{h\nu - E_f}{kT}\right)}\right) \quad (5)$$

where $N_{SS}(E_{CBM} - h\nu)$ is the surface states density at an offset of $h\nu$ below the CBM. Therefore, the Fermi level can be found by fitting the low-energy edge of the surface charge profile using a Fermi function. Since the process of surface charge migration during application of gate stress shifts the Fermi level on the surface to a new position relative to the CBM, the dynamics of the Fermi level as a function of gate stress voltage can be measured and monitored.

**Figure 2** shows a sub-bandgap spectrum of the drain current under illumination at zero gate bias. The current starts to increase at 2.6 eV. The derivative of the photocurrent spectrum is shown in the inset. The fringes observed in both spectra (more pronounced in the derivative) are Fabry-Perot interference fringes caused by internal reflections of the light within the GaN layer. To recover the original surface charge distribution, the effect of the Fabry-Perot interference was cancelled numerically in the subsequent treatment.[17,18,19,16]

**Figure 3** shows a set of surface charge profiles for different gate stress voltages calculated from the photocurrent spectra using Eq. (4). The relative permittivity value used in our calculations was 10.31,[20] and the AlGaN layer thickness – 11 nm. All the profiles have a peak shape and their area decreases as the gate stress voltage is made more negative (depopulating more charge from the surface into the 2DEG). On the other hand, when a positive stress voltage is applied to the gate, the surface charge density increases, above that of the zero gate voltage, due to a partial reverse charge migration from the 2DEG back into the surface states. Overall, the measured surface charge profile appears to be only a small part of a much wider surface states distribution, which has been mostly depleted into the 2DEG.

All the profiles seem to diminish to zero also at their high-energy edge, at ~3.07 eV. However, the positions of their low energy edges are observed to vary gradually with the gate voltage. As discussed previously, this is expected and reflects the change in the surface Fermi level position caused by the change in the density of surface charge with the gate voltage.

The shape of the Fermi function was obtained by a division of $N_{OC}$ by $N_{SS}$ (Eq. (5)). Since the surface density of states, $N_{SS}$, is not known, we used, in its stead, the measured profile at 1 V and extrapolated its right shoulder linearly past the low energy edge. Although this is but a first-order approximation, its validity can be easily tested fitting the result with the Fermi-Dirac distribution at room temperature with a single fitting parameter.

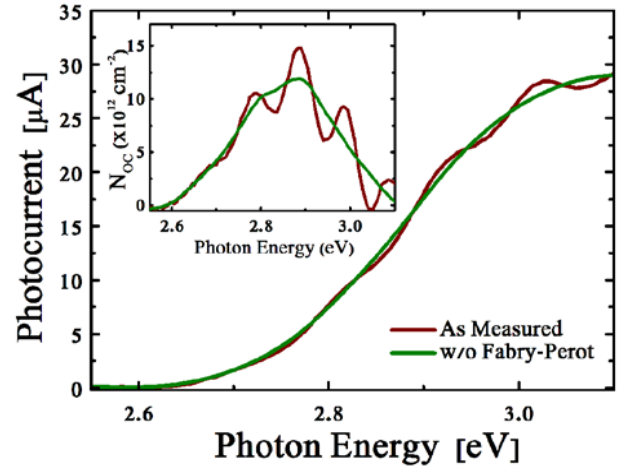

Fig. 2　Photocurrent spectrum obtained at $V_{GS}$=0 V. The inset shows the derivative of the photocurrent spectrum by photon energy. Red lines show the original data, while green lines are the data after cancellation of the Fabry-Perot artifact.

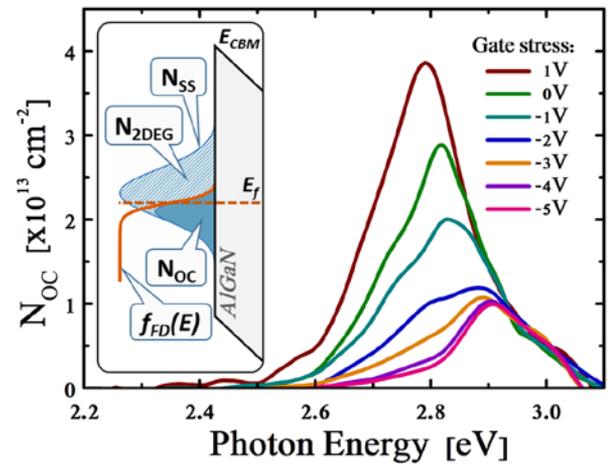

Fig. 3　Surface charge density distribution ($N_{OC}$) profiles as a function of gate stress voltage. Negative gate stress depopulates part of the surface charge into the 2DEG, while positive stress recovers a certain portion of it back to the surface. The inset shows schematically that the density of occupied (charged) surface states ($N_{OC}$) is only a part of the total surface state distribution ($N_{SS}$). The empty part of $N_{SS}$ is the charge that makes up most of the 2DEG sheet charge. Fermi-Dirac distribution function $f_{FD}(E)$ distinguishes between the empty and the occupied surface states.

The resulting Fermi functions are shown in **Fig. 4** (full lines) for a series of gate stress voltage values. The black dashed lines are fits of Fermi-Dirac distribution functions with varying $E_f$. The inset shows the evolution of the Fermi level position as a function of gate voltage (orange dots) acquired from the fit. At zero gate voltage, we found the Fermi level to be positioned 2.8 eV below the CBM. Over the range of our measurements, the position varied from 2.88 eV to 2.74 eV for gate voltages of -5 V and 1 V, respectively. The total surface charge (green triangles in the inset of Fig. 4) was calculated by integration of each of the profiles in Fig. 3, i.e. $n_{OC} = \int N_{OC}(E)dE$. At zero gate bias, the total charge density at the surface was found to be $5.83 \times 10^{12}$ cm$^{-2}$, and its value changed from $8.25 \times 10^{12}$ cm$^{-2}$ to $1.68 \times 10^{12}$ cm$^{-2}$ when the gate voltage was varied from 1 to -5 V respectively. A Fermi level positioned so deep under the CBM implicitly testifies that the origin of the measured charge profiles is at the surface, where the energy difference between the Fermi level and the conduction band reaches its highest value, and not at bulk deep levels.

The proposed method provides invaluable information on the profile of the charged surface state distribution as well as a tool to evaluate the dynamics of the surface Fermi level in actual devices. The method may turn out useful for the surface state engineering that is required in order to control the properties of the 2DEG in nitride HEMTs.

See supplementary material for further detail of the procedure used to obtain the illumination-modified transconductance values, and procedure to remove the Fabry-Perot fringes.

Financial support from the Office of Naval Research Global through a NICOP Research Grant (N62909-18-1-2152) is gratefully acknowledged.

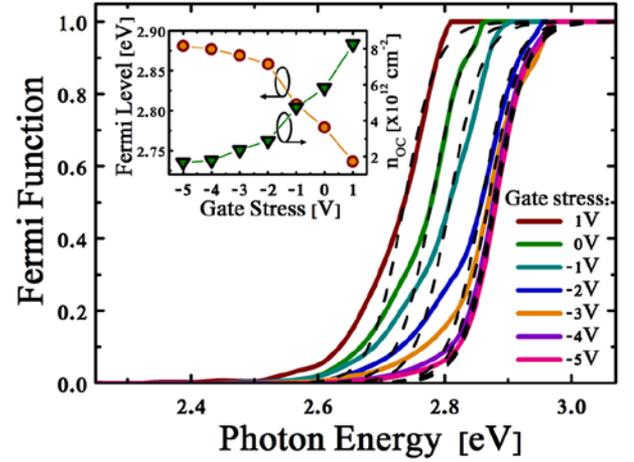

Fig. 4 Fermi functions obtained using Eq. (6) (full lines) along with room temperature Fermi-Dirac distribution fit (dashed lines) as a function of the gate stress voltage. Inset shows the evolution of the calculated Fermi level position and of the total surface charge density, $n_{OC}$, as a function of the gate stress voltage.